\documentclass[journal]{IEEEtran}
\usepackage[T1]{fontenc}
\usepackage{graphicx}
\usepackage{booktabs}
\usepackage{array}
\usepackage{amsmath}
\usepackage{amssymb}
\usepackage{xcolor}
\usepackage{tikz}
\tikzset{
  hum/.style={draw=orange!70!black, fill=yellow!22, rounded corners=2pt, align=center},
  llm/.style={draw=blue!60!black, fill=blue!12, rounded corners=2pt, align=center},
  tool/.style={draw=black!55, fill=black!8, rounded corners=2pt, align=center},
  tag/.style={font=\tiny\bfseries, inner sep=1.2pt, rounded corners=1pt, text=white, anchor=south west, yshift=0.3pt},
  taghum/.style={tag, fill=orange!80!black},
  tagllm/.style={tag, fill=blue!65!black},
  tagtool/.style={tag, fill=black!60},
  arr/.style={->, thick},
  lab/.style={font=\scriptsize, align=center, fill=white, inner sep=1pt}
}
\newcommand{\figlegend}[1]{%
  \node[anchor=center, font=\scriptsize] at #1 {%
    \tikz[baseline=-0.6ex]\node[hum, minimum width=4mm, minimum height=2.5mm]{};~Human\quad
    \tikz[baseline=-0.6ex]\node[llm, minimum width=4mm, minimum height=2.5mm]{};~LLM\quad
    \tikz[baseline=-0.6ex]\node[tool, minimum width=4mm, minimum height=2.5mm]{};~EDA tool};}

\usetikzlibrary{arrows.meta,positioning,calc}
\usepackage[hidelinks]{hyperref}
\newcommand{\VT}{\textit{VibeTapeout}}

\begin{document}
\title{The Era of Vibe Tapeout:\\Dark Age or Golden Age\\for Analog Circuit Design?}
\author{Kentaro Yoshioka%
\thanks{Keio University, Japan.}}
\maketitle

\begin{abstract}
General-purpose LLM coding agents can now take three core mixed-signal building blocks---a SAR ADC, a $\Delta\Sigma$ modulator, and an integer-N PLL---each from a written specification to signoff (a DRC/LVS-clean layout meeting its specifications after parasitic extraction, PEX) on the open SKY130 PDK, with no GUI use and no circuit-specific training.
In this \emph{vibe tapeout} workflow, agents write their own disposable schematic and layout generators, run EDA tools in batch, diagnose post-PEX degradation, and iterate while the human stays outside the loop.
Without a reference design, an 11-bit, 5.12-MS/s SAR ADC reached post-PEX signoff at 62.0\,dB SNDR; given a sanitized reference design (\emph{vibe porting}), it reached 63.7\,dB in 0.036\,mm$^2$ and 18.9\,h with no human intervention.
An integer-N PLL reached signoff from specification alone (0.28\,mm$^2$, 9.8-ps integrated jitter including device noise).
So did a third-order $\Delta\Sigma$ modulator (0.977\,mm$^2$, 96.6\,dB SNDR, 4.35\,mW, 163.2\,dB Schreier FoM).
In the SAR ADC, vibe porting cuts layout area to 1/1.6--1/2.8 of from-scratch runs, and agents diagnose subtle post-PEX mechanisms, such as wiring coupling in the capacitive DAC (CDAC), through their own controlled experiments.
Conversely, an unsupervised agent using parallel subagents laid out a 13-bit SAR ADC in 28.6\,min yet missed its own post-PEX target; because agents sometimes report false passes, verification independent of the design agent is essential.
Because commercial PDKs cannot be sent to cloud LLMs, we also evaluate local LLMs on a six-task benchmark on a commercial 65-nm CMOS PDK: Kimi K3 passed five tasks and MiniMax M3 two, while neither solved symptom-based debugging.
Finally, we ask whether this era is a dark age or a golden age for analog design.
Our answer is that it is a golden age for designers who know what to verify.
\end{abstract}

\begin{IEEEkeywords}
Large language models, LLM agents, analog circuit design, SAR ADC, PLL, delta-sigma modulator, benchmark, open PDK, local LLM
\end{IEEEkeywords}

\section{Introduction}
Give an unmodified, general-purpose LLM coding agent a written specification, command-line access to standard EDA tools, and a day or two of compute, and it can now take the core mixed-signal building blocks of a chip all the way through layout to signoff.
On the open SkyWater 130-nm PDK (SKY130), LLM agents carried three key blocks---an 11-bit SAR ADC, a third-order $\Delta\Sigma$ modulator, and an integer-N PLL---each from written specification to DRC/LVS-clean signoff after parasitic extraction (PEX) without opening a GUI, without circuit-specific training, and with only a few high-level human prompts (Fig.~\ref{fig:gallery}).
Just as ``vibe coding''\cite{karpathy} lets software engineers steer LLMs through high-level intent and test outputs without reading every line of code, integrated-circuit design is entering the era of \emph{vibe tapeout}: agents write their own schematic and layout generators, run circuit simulation, DRC, LVS, and parasitic extraction in batch, interpret the results, and revise the design while the human stays outside the loop.

\begin{figure*}[t]\centering
\input{figs/gallery.tex}
\caption{Four mixed-signal building blocks taken from specification to post-PEX signoff on SKY130 by LLM agents. All four layouts are rendered from GDS with the same layer colors; each panel has its own scale bar, and boxes mark main blocks (BSW: bootstrapped sampling switch; DWA: data-weighted averaging; CIFF: cascade of integrators with feedforward; POR: power-on reset; MIM: metal-insulator-metal; CP: charge pump; PFD: phase-frequency detector; AAF: anti-aliasing filter; PGA: programmable-gain amplifier; LNA: low-noise amplifier). Agents: (a) Claude Sonnet 5.5 with sanitized reference (VibePorting); (b) GPT-6 (Codex), spec + Verilog-A model; (c) GPT-6.1 Sol (Codex), spec only; (d) Gemini 4 Neon, spec only. See Table~\ref{tab:blocks}.}
\label{fig:gallery}
\end{figure*}

While LLM-assisted RTL generation is advancing rapidly in digital design, analog circuits do not end at the schematic.
Device placement, symmetry, and femtofarad-scale wiring parasitics determine whether silicon works.
For analog designers, the question is therefore not whether an LLM can write a netlist, but whether it can carry a block through layout and post-layout verification. This paper addresses three research questions:
\begin{itemize}\IEEEsetlabelwidth{\textbf{RQ3}}
\item[\textbf{RQ1}] \textbf{Breadth:} Can general-purpose LLM agents take the key mixed-signal building blocks---a SAR ADC, a $\Delta\Sigma$ modulator and a PLL---from a written specification to signoff, including layout and post-layout verification?
\item[\textbf{RQ2}] \textbf{Depth:} What post-layout quality can be expected and what drives it---a reference design, human intervention---and where do agents fail after layout, can they diagnose it themselves, and how must verification change?
\item[\textbf{RQ3}] \textbf{Behind an NDA:} Can the same approach be used with commercial PDKs, which cannot be exposed to cloud LLMs?
\end{itemize}

After background (Section~\ref{sec:bg}), we answer these questions in three steps.
For RQ1, on SKY130, where cloud LLM agents can be operated freely, we carry all three blocks to post-PEX signoff with one common framework, which we call \VT{} (Sections~\ref{sec:method} and~\ref{sec:blocks}).
For RQ2, we examine the SAR ADC in depth (Section~\ref{sec:sar}): porting from a sanitized reference design (\emph{vibe porting}; condition VibePorting) versus design from scratch (conditions FromScratch and FromScratch-VA), human interventions, autonomous post-PEX diagnosis, and a 13-bit SAR ADC laid out by parallel subagents.
For RQ3, because commercial PDK files cannot be exposed to cloud LLMs, we evaluate local open-weight LLMs on a six-task benchmark on a commercial 65-nm CMOS PDK (Section~\ref{sec:local}).
Section~\ref{sec:howto} distills what designers can delegate today, Section~\ref{sec:disc} discusses why this is possible now and whether this is a dark or a golden age for analog design, and Section~\ref{sec:conc} concludes.

\section{Background}
\label{sec:bg}
\subsection{The Generator Approach}
Efforts to make IC design faster and easier long predate LLMs.
At their center is the idea of the \emph{generator}, pioneered by M.~Horowitz and colleagues at Stanford\cite{cmpgen,chipgen}.
Its message is: ``do not design a chip; design a generator that produces chips.''
If the designer's knowledge and trade-offs are captured in a parameterized generator, chips optimized for each application can be produced repeatedly just by changing the settings.
The idea evolved into the AHA project\cite{aha}, which aims at a development flow in which hardware can be rebuilt again and again, like software.
In analog design, BAG\cite{bag,bag2} describes the designer's sizing and layout procedures in Python so that they can be re-run across processes.
FASoC\cite{fasoc} generates analog blocks as compositions of synthesizable cells,
and ALIGN\cite{align} and MAGICAL\cite{magical} automate placement and routing by extracting constraints from the netlist.
For SAR ADCs, a BAG-based generator for time-interleaved SAR ADCs\cite{tisar_bag}, the template-and-grid layout generation engine LAYGO\cite{laygo}, and OpenSAR\cite{opensar}, which generates a DRC/LVS-clean layout from performance specifications, have been reported.

These works also showed that building the generator itself becomes the central design task.
The Chip Generator paper notes that constructing a useful generator is quite challenging\cite{chipgen}.
Writing a good generator requires both deep circuit and layout knowledge and skilled programming.
The investment pays off by reusing the same generator for many instances (Fig.~\ref{fig:loop}(a)).

LLM agents are changing who does this job of writing generators.
The agents in this work write layout generators, check scripts and evaluation pipelines for each block on the spot,
and repeatedly rewrite them after reading the results (Fig.~\ref{fig:loop}(b)).
When writing a generator becomes cheap, \emph{disposable generators} that do not assume reuse become an option.
This does not replace conventional generator research; rather, it pushes the ``generate, don't design'' direction from a different angle.

\subsection{Circuit Design with LLMs}
Applications of LLMs to circuit design began with Verilog generation\cite{chipchat,verilogeval} and domain-adapted chat assistants\cite{chipnemo}.
For analog circuits, netlist generation\cite{analogcoder} and interactive layout\cite{layoutcopilot} have been reported.
Most of these use the LLM as a ``component'' for a specific step.
Recently, methods that constrain LLM agents with expert templates for SAR ADCs\cite{atlas} and public projects in which agents design autonomously\cite{aisar,gfsar} have also appeared (compared in Section~\ref{sec:sar_conditions}).
In contrast, this paper studies general-purpose coding agents that operate the entire design flow through the shell.
The agents write the generators, the check scripts and the evaluation pipelines on the spot.
Little tool-specific preparation is required.
This is the biggest difference from conventional automation.

\begin{figure*}[t]\centering\color{black}
\begin{minipage}[b]{0.47\textwidth}\centering
\resizebox{\linewidth}{!}{
\begin{tikzpicture}[font=\footnotesize, >=Stealth]
\node[hum, dashed, text width=78mm, align=left, inner sep=3pt, anchor=south] (exp) at (0, 3.15)
  {\textbf{Circuit/layout experts}: encode design procedures and\\
   know-how as a generator (domain-specific language / framework)};
\node[taghum] at (exp.north west) {Human};
\node[hum, minimum width=60mm, minimum height=11mm] (gen) at (0, 2.15)
  {\textbf{Hand-write a general generator}\\ large up-front investment per circuit class and process};
\node[taghum] at (gen.north west) {Human};
\draw[arr] (exp.south) -- (gen.north);
\node[tool, minimum width=20mm, minimum height=11mm] (i1) at (-2.6, 0.1) {Instance 1\\{\scriptsize (spec A)}};
\node[tool, minimum width=20mm, minimum height=11mm] (i2) at (0, 0.1)    {Instance 2\\{\scriptsize (spec B)}};
\node[tool, minimum width=20mm, minimum height=11mm] (i3) at (2.6, 0.1)  {Instance $N$\\{\scriptsize (new process)}};
\node[tagtool] at (i1.north west) {generated};
\foreach \i in {i1,i2,i3} \draw[arr] (gen.south) -- node[lab, pos=0.5] {\scriptsize params} (\i.north);
\node[font=\scriptsize] at (1.3,0.1) {$\cdots$};
\node[hum, minimum width=60mm, minimum height=11mm] (ver) at (0,-1.75)
  {\textbf{Humans run the tools and judge the results}\\ bugs and unforeseen specs go back into the generator};
\node[taghum] at (ver.north west) {Human};
\foreach \i in {i1,i2,i3} \draw[arr] (\i.south) -- (ver.north);
\draw[arr, dashed, orange!70!black] (ver.west) to[bend left=50] node[lab, left=1pt, pos=0.5, text=orange!70!black] {fix\\(manual)} (gen.west);
\node[draw, dashed, rounded corners=2pt, text width=60mm, align=center, anchor=north] (note) at (0,-2.9)
  {\textbf{Value}: write once, use many times;\\the investment is recovered by \emph{reuse}};
\figlegend{(0,-5.75)}
\end{tikzpicture}}\\[1mm]
{\small (a) Conventional: hand-written general generator}
\end{minipage}\hfill
\begin{minipage}[b]{0.47\textwidth}\centering
\resizebox{\linewidth}{!}{
\begin{tikzpicture}[font=\footnotesize, >=Stealth]
\node[llm,  minimum width=33mm, minimum height=11mm] (gen) at (0, 2.15)  {\textbf{(1) Write / fix generator}\\ Python $\rightarrow$ SPICE netlist\\ Python/Tcl $\rightarrow$ Magic layout};
\node[tagllm] at (gen.north west) {LLM};
\node[tool, minimum width=33mm, minimum height=11mm] (run) at (3.15, 0)  {\textbf{(2) Run tools in batch}\\ SPICE simulation\\ Magic DRC/PEX, Netgen LVS};
\node[tagtool] at (run.north west) {Tool};
\node[llm,  minimum width=33mm, minimum height=11mm] (res) at (0, -2.15) {\textbf{(3) Turn results into text}\\ metrics (SNDR, jitter, \ldots)\\ DRC count, LVS verdict\\ extracted R and C, layout PNG};
\node[tagllm] at (res.north west) {LLM-written scripts};
\node[llm,  minimum width=33mm, minimum height=11mm] (rsn) at (-3.15, 0) {\textbf{(4) Read and diagnose}\\ form a hypothesis, plan\\ one-change experiments};
\node[tagllm] at (rsn.north west) {LLM};
\node[draw=blue!60!black, very thick, circle, align=center, fill=blue!12, minimum size=17mm] (agent) at (0,0) {LLM\\agent};
\draw[arr, blue!60!black] (gen.east) to[bend left=22] node[lab, pos=0.55, above right=-1pt] {run} (run.north);
\draw[arr] (run.south) to[bend left=22] node[lab, pos=0.45, below right=-1pt] {output} (res.east);
\draw[arr, blue!60!black] (res.west) to[bend left=22] node[lab, pos=0.55, below left=-1pt] {read} (rsn.south);
\draw[arr, blue!60!black] (rsn.north) to[bend left=22] node[lab, pos=0.45, above left=-1pt] {fix} (gen.west);
\foreach \n in {gen,run,res,rsn} \draw[gray, densely dotted] (agent) -- (\n);
\node[hum, dashed, text width=86mm, align=left, inner sep=3pt, anchor=south] (hum) at (0, 3.15)
  {\textbf{Human designer} (outside the loop): spec, acceptance criteria and policy;\\
   milestone reviews (layout images, post-PEX loss); hints if needed};
\node[taghum] at (hum.north west) {Human};
\draw[arr, dashed, orange!70!black] (hum.south) -- node[lab, right=1pt, text=orange!70!black] {goals, hints} (gen.north);
\node[hum, dashed, anchor=north, text width=24mm, align=left, inner sep=3pt] (ref) at (-3.15, -2.9)
  {\textbf{Reference package}\\(optional)\\architecture, relative\\sizes, design notes};
\node[taghum] at (ref.north west) {Human-prepared};
\draw[arr, dashed, orange!70!black] (ref.north) -- (rsn.south west);
\node[draw, dashed, rounded corners=2pt, anchor=north, text width=27mm, align=left, inner sep=3pt] (out) at (3.15, -2.9)
  {\textbf{Artifacts}\\ GDS, netlists,\\ structured design\\ and iteration logs};
\draw[arr, dashed] (res.south east) -- (out.north);
\begin{scope}[shift={(0,-5.75)}]
\foreach \x/\t [count=\i] in {-3.6/{Functional\\{\scriptsize (Verilog-A)}}, -1.8/{Schematic}, 0/{Layout}, 1.8/{PEX}, 3.6/{Signoff\\{\scriptsize checks}}}
  \node[llm, minimum width=15mm, minimum height=8.5mm, font=\scriptsize] (s\i) at (\x,0) {\t};
\foreach \a/\b in {1/2,2/3,3/4,4/5} \draw[arr] (s\a.east) -- (s\b.west);
\node[font=\scriptsize, align=center] at (0,0.68) {At every stage the LLM runs loop (1)--(4) and moves on once that stage's acceptance criteria are met};
\end{scope}
\figlegend{(0,-6.7)}
\end{tikzpicture}}\\[1mm]
{\small (b) Vibe tapeout: LLM-written generators and the design loop}
\end{minipage}
\caption{Two forms of generator-based design. Colors indicate who does the work (yellow: human, blue: LLM, gray: EDA tools). (a) Experts codify design knowledge by hand as a general generator and recover the investment through reuse. (b) An LLM agent writes generators for each task, runs EDA tools in batch, and reflects the results back into the generators; this loop runs at every stage from the functional model to signoff. The human stays outside the loop.}
\label{fig:loop}
\end{figure*}

\section{How the Agents Work}
\label{sec:method}
All three SKY130 building blocks in Section~\ref{sec:blocks} and the controlled SAR study in Section~\ref{sec:sar} are run with the same framework, \VT{}: the proper name of this paper's specific design loop, tool environment, milestone structure (M1--M5), standard checkpoints (CP-A/B/C), and independent verification protocol, as distinct from \emph{vibe tapeout}, the general practice of letting LLM agents take a design to tapeout.

\subsection{The Autonomous Design Loop and Tool Environment}
As shown in Fig.~\ref{fig:loop}(b), an agent writes Python or Tcl programs that emit netlists and layouts, invokes circuit simulation, DRC, LVS, and parasitic extraction in batch, and inspects text summaries (SNDR, jitter, DRC violation counts, LVS verdicts, extracted R and C) together with rendered layout PNGs.
From these observations it forms a hypothesis, edits its generators, and reruns the tools, iterating tens to hundreds of times with each cycle taking a few minutes to tens of minutes.

All open-PDK runs target the SkyWater SKY130 PDK\cite{sky130}, using its 1.8-V devices and the \texttt{sky130\_fd\_sc\_hd} standard cells.
Layout and parasitic extraction use Magic, LVS uses Netgen, and circuit simulation uses a SPICE simulator with the public SKY130 models, all accessed via thin shell wrapper scripts returning concise text outputs.
The scope ends at macro-level PEX (no pads) and excludes Monte Carlo device-mismatch simulation.

\subsection{Roles: Human, Supervising Agent, and Design Agents}
To keep prompts uniform and prevent reliance on agent self-reports, \VT{} separates three actors (Fig.~\ref{fig:framework}).

\begin{figure*}[t]\centering\color{black}
\resizebox{0.76\textwidth}{!}{
\begin{tikzpicture}[font=\footnotesize, >=Stealth]
\node[hum, text width=34mm, align=left, inner sep=3pt] (hum) at (-8.25, 1.2)
  {\textbf{Human designer}\\spec and acceptance criteria\\policy (run control,\\\hspace*{1em}area caps, deadlines)\\layout guideline, hints};
\node[taghum] at (hum.north west) {Human};
\node[llm, text width=47mm, align=left, inner sep=3pt] (sup) at (-1.75, 1.2)
  {\textbf{Supervising agent}\\[1pt]
   $\bullet$ writes prompts and the protocol\\
   $\bullet$ standard checkpoints CP-A/B/C\\
   $\bullet$ monitors progress, sets deadlines\\
   $\bullet$ escalates when an agent stalls\\
   $\bullet$ \textbf{independent verification}:\\\hspace*{0.8em}metrics (SNDR, jitter, \ldots)\\\hspace*{0.8em}from raw waveforms; DRC/LVS rerun\\
   $\bullet$ isolation audit\\
   $\bullet$ logs every intervention};
\node[tagllm] at (sup.north west) {LLM};
\draw[arr, orange!70!black] ($(hum.east)+(0,0.3)$) -- node[lab, above, text=orange!70!black, font=\scriptsize] {decisions} ($(sup.west)+(0,0.3)$);
\draw[arr] ($(sup.west)+(0,-0.3)$) -- node[lab, below, font=\scriptsize] {reports, questions} ($(hum.east)+(0,-0.3)$);
\foreach \y/\n/\t in {2.6/ref/{Run 1: SAR ADC}, 1.2/bl/{Run 2: $\Delta\Sigma$}, -0.2/va/{Run $N$: PLL, \ldots}} {
  \node[draw=gray, dashed, rounded corners=3pt, minimum width=42mm, minimum height=11.5mm] (box\n) at (4.75, \y) {};
  \node[llm, text width=23mm, align=center, inner sep=2pt] (d\n) at (4.05, \y) {\textbf{Design agent}\\\t};
}
\node[tagllm] at (dref.north west) {LLM};
\node[hum, dashed, text width=11.5mm, align=center, inner sep=1pt, font=\scriptsize] (pkg) at (6.02, 2.6) {Reference\\package\\(optional)};
\node[font=\scriptsize, gray, anchor=south east] at (boxref.north east) {sandbox (own work directory only)};
\node[font=\scriptsize, align=left, anchor=west] at (5.45, 1.2) {spec};
\node[font=\scriptsize, align=left, anchor=west] at (5.45, -0.2) {spec\\(+ jitter\\method)};
\foreach \n in {ref, bl, va} \draw[arr, blue!60!black] ($(sup.east)+(0,0.25)$) -- (box\n.west);
\node[lab, text=blue!60!black, align=center, font=\scriptsize] at (1.75, 3.05) {prompts\\CP-A/B/C};
\coordinate (ra) at ($(boxva.west)+(0,-0.35)$); \draw[arr, densely dashed] (ra) -- node[lab, below, font=\scriptsize, align=center] {artifacts, logs} (sup.east |- ra);
\node[llm, dashed, text width=26mm, align=center, inner sep=2pt] (help) at (2.2, -1.85) {Helper agent\\(any run, if stalled)};
\node[tagllm] at (help.north west) {LLM};
\draw[arr, blue!60!black, dashed] ($(sup.south east)+(-0.4,0)$) |- (help.west);
\draw[arr, blue!60!black, dashed] (help.east) -| node[lab, pos=0.75, right, font=\scriptsize] {diagnose, fix} ($(boxva.south)+(-0.6,0)$);
\node[tool, align=center, inner sep=4pt, minimum width=165mm] (tool) at (-1.5, -3.35)
  {\textbf{EDA tools and compute}: Magic (layout, DRC, PEX), Netgen (LVS), SPICE simulator, Yosys (logic synthesis), batch compute};
\node[tagtool] at (tool.north west) {Tool};
\coordinate (lp) at ($(boxva.south)+(1.2,0)$); \draw[arr] (lp) -- node[lab, right, font=\scriptsize] {design loop (Fig.~\ref{fig:loop}(b))} (lp |- tool.north);
\coordinate (vp) at ($(sup.south)+(-1.0,0)$); \draw[arr, blue!60!black] (vp) -- node[lab, left, font=\scriptsize] {independent verification} (vp |- tool.north);
\figlegend{(-1.5,-4.3)}
\end{tikzpicture}}
\caption{The \VT{} framework, common to all supervised SKY130 runs. The human sets the policy and specification, and the supervising agent runs the experiment. Each run (one block, or one SAR condition in Section~\ref{sec:sar}) has its own design agent, isolated in a sandbox and executing the design loop (Fig.~\ref{fig:loop}(b)); a reference package is optional. The supervising agent issues identical checkpoint texts and independently verifies all reported deliverables.}
\label{fig:framework}
\end{figure*}

\begin{table*}[t]
\centering
\caption{Four mixed-signal building blocks taken to post-PEX signoff on SKY130. Each column has its own specification and targets; SAR shows VibePorting, $\Delta\Sigma$ used spec + Verilog-A model, and PLL and AFE used spec alone. Times are wall-clock; performance is post-PEX at TT (27\,$^\circ$C). TIE: time-interval error.}
\label{tab:blocks}
\scriptsize
\setlength{\tabcolsep}{2.5pt}
\renewcommand{\arraystretch}{1.08}
\begin{tabular}{@{}>{\raggedright\arraybackslash}p{0.13\textwidth} >{\raggedright\arraybackslash}p{0.205\textwidth} >{\raggedright\arraybackslash}p{0.205\textwidth} >{\raggedright\arraybackslash}p{0.205\textwidth} >{\raggedright\arraybackslash}p{0.205\textwidth}@{}}
\toprule
 & \textbf{11-bit SAR ADC} (Sec.~\ref{sec:sar_headline}) & \textbf{3rd-order $\Delta\Sigma$ modulator} (Sec.~\ref{sec:dsm}) & \textbf{Integer-N PLL} (Sec.~\ref{sec:pll}) & \textbf{Audio mic AFE} (Sec.~\ref{sec:afe}) \\
\midrule
Design agent & Claude Sonnet 5.5; supervised, CP-A/B/C & GPT-6 (Codex); supervised, CP-A/B/C & GPT-6.1 Sol (Codex); supervised, CP-A/B/C & Gemini 4 Neon \\
Specification \& reference & Scaled from our 65-nm design + sanitized reference (VibePorting) & Scaled from our 65-nm design; spec + Verilog-A behavioral model & Written by supervising agent; spec only & Written for our $\Delta\Sigma$ load; spec only \\
Human inputs & Common fixed inputs only (Sec.~\ref{sec:sar_conditions}); 0 interventions & 5 high-level prompts (incl.\ milestone transitions and a $\leq$1-mm$^2$ area cap) & 3 high-level prompts (incl.\ a $\leq$0.3-mm$^2$ area cap) & -- \\
Time: schematic (M2) / layout / signoff (M5) & 4.2\,h / 0.9\,h$^{a}$ / 18.9\,h & -- / -- / -- & 33.7\,h / 44.2\,h / 52.4\,h & -- / -- / -- \\
Area [mm$^2$], DRC/LVS & 0.036; clean / clean & 0.977; clean / clean & 0.279; clean / clean & 0.991; clean / clean \\
PEX; noise & RC; noiseless & RC; noiseless & RC; model-based noise estimate & PEX; device noise \\
Post-PEX result & SNDR 63.71\,dB, SFDR 76.17\,dB (256-pt FFT); 307.8\,$\mu$W (SS/FF: 64.74/64.22\,dB) & SNDR 96.6\,dB, HD3 $-101.4$\,dBc (2048-pt Hann FFT, $f_\mathrm{in}=6$\,kHz); 4.35\,mW; Schreier FoM 163.2\,dB & Cold-start lock 7.8\,$\mu$s; jitter 9.8\,ps (TIE + device noise); spur $-44.8$\,dBc; 4.60\,mW & Gain 6--24\,dB, $f_{-3\mathrm{dB}}=25.1$\,kHz; IRN 2.35\,$\mu$V$_\mathrm{rms}$(A), DR 100.6\,dB(A); THD $-84.6$/$-78.9$\,dB (CT/SC); CMRR/PSRR 95.4/105.8\,dB; 1.88\,mW \\
Status & \textbf{Signoff (TT/SS/FF)} & \textbf{Signoff (TT)} & \textbf{Signoff (TT)} & \textbf{Signoff (TT)} \\
\bottomrule
\multicolumn{5}{@{}p{0.98\textwidth}@{}}{\scriptsize $^{a}$From the start of layout to the first DRC/LVS-clean layout.}
\end{tabular}
\end{table*}

\begin{itemize}
\item \textbf{Design agent:} A general-purpose coding agent---Claude Sonnet 5.5\cite{sonnet55} for the 11-bit SAR ADC runs, GPT-6\cite{gpt6} for the $\Delta\Sigma$ modulator, and GPT-6.1 Sol\cite{gpt61sol} for the integer-N PLL (the latter two with Codex\cite{codex}); Section~\ref{sec:sar13b} also examines an unsupervised run in which Gemini 4 Neon\cite{gemini4} received only two brief prompts. Each supervised design agent runs in an isolated sandbox with read access only to its own working directory and shared tool wrappers. Copying existing open-source designs is forbidden and every web access is logged. Inside the sandbox, the agent carries the block from functional model to signoff while recording its reasoning, state, and handoff notes in structured files.
\item \textbf{Supervising agent:} A separate LLM agent that administers the run: it sends prompts and standardized checkpoints with identical wording, tracks progress, inspects tool-call logs to audit isolation, and performs \emph{independent verification}. It gives no design advice. When a design agent stalls, the supervising agent may escalate to a \emph{helper agent} running a more capable model (e.g., Claude Opus 5.5\cite{opus55}) inside the same isolated directory; any helper call is logged as an intervention.
\item \textbf{Human designer:} Sets specifications, acceptance criteria, and high-level policy (such as area caps or deadlines), and reviews milestone escalations reported by the supervising agent while remaining outside the iteration loop.
\end{itemize}

\subsection{Milestones, Checkpoints, and Independent Verification}
Every supervised block advances through five milestones: M1 (functional model), M2 (schematic meeting specification at TT and functioning at SS/FF), M3 (DRC/LVS-clean layout), M4 (post-PEX simulation), and M5 (final signoff report).
The supervising agent delivers three identical standard checkpoints---after M3 (CP-A), after M4 (CP-B), and before M5 signoff (CP-C); the exact texts are given in Appendix~\ref{app:cp}:
\textbf{CP-A} instructs the agent to inspect a full-scale PNG rendering of its layout and improve empty area, differential symmetry, and sensitive-net routing;
\textbf{CP-B} instructs it to explain the schematic-to-PEX gap via one-factor-at-a-time experiments and fix the largest contributor;
and \textbf{CP-C} asks it to list and check failure modes invisible to DRC and LVS.

Crucially, the supervising agent never accepts a design agent's reported metrics at face value.
An independent verifier re-runs DRC and LVS against the simulated schematics and recomputes headline performance metrics (such as FFT-based SNDR) directly from raw simulator waveforms.
In what follows, all SAR ADC and PLL numbers have passed this independent re-verification.

\section{Building Blocks by Vibe Tapeout}
\label{sec:blocks}
Table~\ref{tab:blocks} summarizes the four blocks; each targets its own circuit class and specification, so the columns are not repetitions of one task.

\subsection{11-Bit SAR ADC}
\label{sec:sar_headline}
The SAR ADC specification was obtained by scaling our 65-nm CMOS design to SKY130 with a 1.8-V supply: 11-bit resolution, 5.12\,MS/s sampling rate, post-PEX TT targets of SNDR $\geq$ 60\,dB and spurious-free dynamic range (SFDR) $\geq$ 70\,dB (256-point coherent FFT, $-1$\,dBFS), power $\leq$ 450\,$\mu$W, and area $\leq$ 0.14\,mm$^2$.
In our primary reference-assisted run (\emph{VibePorting}), Claude Sonnet 5.5 received a sanitized reference package derived from that design plus the common fixed inputs (both defined in Section~\ref{sec:sar_conditions}), and reached post-PEX signoff at TT, SS, and FF with zero human interventions and no helper agent (Table~\ref{tab:blocks}, Fig.~\ref{fig:gallery}(a)).
Section~\ref{sec:sar} compares it with from-scratch runs and traces how the agent recovered from its first post-PEX drop.

\subsection{Third-Order $\Delta\Sigma$ Modulator}
\label{sec:dsm}
For a high-linearity, noise-limited block, GPT-6 (Codex) designed a third-order $\Delta\Sigma$ modulator from its specification and a Verilog-A behavioral model, scaled from our 65-nm design: $f_s = 6.144$\,MHz, oversampling ratio (OSR) 128, post-PEX SNDR $\geq$ 95\,dB in the 20-kHz band \emph{with noise}, power $\leq$ 5\,mW, and Schreier FoM $\geq$ 160\,dB.
The agent built a fully differential switched-capacitor modulator with a cascade-of-integrators feedforward (CIFF) loop, a 4-bit quantizer, and data-weighted averaging (DWA) (98.1\,dB SNDR with noise at 3.41\,mW in the schematic) and laid it out under a 1-mm$^2$ area cap (Fig.~\ref{fig:gallery}(b)).
Along the way it traced a post-PEX loss to quantizer output lines coupling into the bias nodes of the first-integrator amplifier and rerouted them.
With RC extraction it met every TT specification (96.6\,dB SNDR, 4.35\,mW, 163.2\,dB FoM).

\subsection{Integer-N PLL}
\label{sec:pll}
For a phase-domain block, GPT-6.1 Sol (Codex) designed an integer-N charge-pump PLL from specification alone: 0.4--1.0\,GHz ($N = 16$--$40$), lock within 20\,$\mu$s, integrated jitter $\leq$ 10\,ps \emph{including device noise}, spur $\leq -40$\,dBc, and power $\leq$ 5\,mW post-PEX, with the loop filter integrated within the extracted PLL footprint.
Because the open-source simulator used for the PLL (ngspice) has no periodic-noise analysis, the specification prescribed how to evaluate jitter: extracted block noise combined in a phase-domain loop model, plus a long noiseless transient for deterministic jitter and spurs.
Under a 0.3-mm$^2$ area cap, the agent built a ring-VCO PLL with an integrated metal-insulator-metal (MIM) loop filter (Fig.~\ref{fig:gallery}(c)) that met every TT target with RC extraction at $N=32$ (Table~\ref{tab:blocks}) and divides correctly over $N=16$--$40$.
Notably, the agent audited itself: it retracted its own milestones when it found an error in its noise projection, and caught a false lock indication at $N=23$ on a revision that had already passed earlier milestones.

\subsection{Audio Microphone AFE}
\label{sec:afe}
For a continuous-time, low-noise analog block, Gemini 4 Neon designed a fully differential audio microphone AFE from specification alone to drive our $\Delta\Sigma$ modulator's switched-capacitor load ($C_s = 23.8$\,pF/side at 6.144\,MHz): programmable gain 6--24\,dB, $f_{-3\mathrm{dB}} = 22$--26\,kHz, IRN $\leq 2.5$\,$\mu$V$_\mathrm{rms}$(A), DR $\geq 100$\,dB(A), THD $\leq -75$\,dB ($-1$\,dBFS), CMRR $\geq 70$\,dB, PSRR $\geq 60$\,dB, and power $\leq 2.0$\,mW.
The agent built a capacitively coupled PGA with local-well pseudo-resistors, a second-order MFB active-RC Butterworth AAF with a split-feedback kickback filter, and a bias/VCM buffer in a mirror-symmetric layout (0.991\,mm$^2$, DRC/LVS clean; Fig.~\ref{fig:gallery}(d)).
Post-PEX review caught virtual-ground capacitive asymmetry and high-temperature leakage offset, prompting the mirror-symmetric PGA and balanced switches; post-PEX at TT, the AFE met its targets (Table~\ref{tab:blocks}), with 105.2\,dB alias rejection at 6.12\,MHz and 6-kHz THD of $-84.6$\,dB (CT) and $-78.9$\,dB (SC $\Delta\Sigma$ load).

\begin{table*}[t]\centering
\caption{SKY130 11-bit SAR ADC under three conditions, all by Claude Sonnet 5.5 (TT, 27\,$^\circ$C, 256-point FFT, $-1$\,dBFS). Wall-clock times; first-layout time counts from the start of layout. Interventions exclude the common fixed inputs (Section~\ref{sec:sar_conditions}). $\ddagger$: at SS, conversions did not complete within the allotted conversion time (spectrum invalid).}
\label{tab:sar}
\footnotesize
\begin{tabular}{lccc}
\toprule
 & VibePorting & FromScratch & FromScratch-VA \\
\midrule
Schematic SNDR/SFDR [dB] & 65.99/79.14 & 61.21/73.35 & 66.79/80.83 \\
Schematic power [$\mu$W] & 204.5 & 346.6 & 167.2 \\
First PEX SNDR/SFDR [dB] & 57.67/66.31 & 52.19/66.28 & 53.87/67.92 \\
Final PEX SNDR/SFDR [dB] & \textbf{63.71/76.17} & 59.90/70.53 (fail) & 61.95/75.40 \\
PEX SS / FF SNDR [dB] & 64.74 / 64.22 & $\ddagger$ / 55.42 (fail) & 60.14 / 63.26 \\
Final PEX power [$\mu$W] & 307.8 & 384.1 & 307.7 \\
Area [$\mu$m$^2$] & \textbf{35{,}723} & 99{,}745 & 55{,}498 \\
Time to schematic (M2) & 4.2\,h & 5.1\,h & 2.2\,h \\
Time to first layout (M3) & \textbf{0.9\,h} & 8.9\,h & 11.3\,h \\
Time to signoff (M5) & \textbf{18.9\,h} & not reached & 26.8\,h \\
Interventions & \textbf{none} & 4 human hints + helper & helper agent \\
\bottomrule
\end{tabular}
\end{table*}

\begin{figure*}[t]\centering
\includegraphics[width=0.92\textwidth]{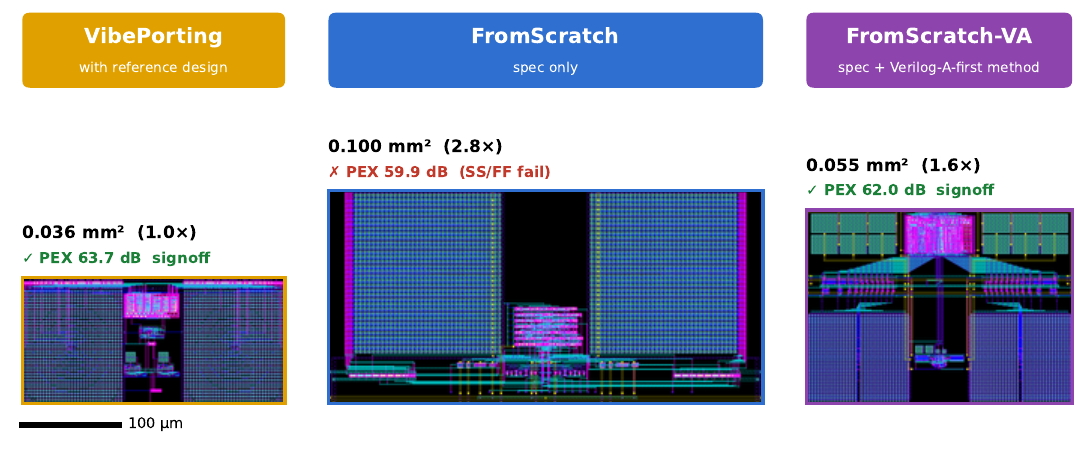}
\caption{Layouts of the SKY130 11-bit SAR ADCs (all at the same scale; Claude Sonnet 5.5 for all three conditions). Above each layout are its area (ratio to VibePorting) and post-PEX SNDR at TT ($\checkmark$: specification met). VibePorting ported the reference floorplan and reached signoff first with the smallest area.}
\label{fig:sarlayout}
\end{figure*}

\subsection{What Each Block Teaches}
\label{sec:teaches}
These blocks span distinct regimes---charge-redistribution switching and asynchronous timing (SAR ADC), noise-limited precision ($\Delta\Sigma$ modulator), phase-domain loop dynamics (PLL), and continuous-time low-noise amplification (AFE)---and four cross-cutting lessons stand out.
First, the same \VT{} loop and checkpoints worked across Claude and GPT agents, and the main post-PEX failure was shared: wiring coupling into sensitive analog nodes.
Second, without an explicit area target, agents produced heavily oversized layouts---by roughly 5$\times$ on the PLL (1.50 to 0.279\,mm$^2$).
Third, each block exposes a different bottleneck: matching and floorplanning dominate the SAR ADC; RC transient verification, rather than layout generation, dominates the high-resolution $\Delta\Sigma$ modulator; and evaluation methodology dominates the PLL, because metrics that the simulator cannot compute directly (such as jitter with device noise) do not carry over automatically and must be prescribed in the specification---introducing modeling assumptions that an independent verifier must audit.
Fourth, under strict milestone definitions an agent can retract its own premature passes (PLL).
With these blocks at signoff, we next dissect the SAR ADC to see \emph{why} it works.

\section{A Closer Look at the SAR ADC}
\label{sec:sar}

\subsection{Vibe Porting Versus From-Scratch Design}
\label{sec:sar_conditions}
To measure the value of existing design assets without violating our commercial PDK's NDA, we ran the 11-bit, 5.12-MS/s SKY130 SAR ADC under three controlled conditions, each executed by Claude Sonnet 5.5 within \VT{} (Fig.~\ref{fig:framework}):
\begin{itemize}
\item \textbf{VibePorting:} Receives a sanitized reference package derived from our 65-nm design: the architecture, a netlist with generalized device names and $W/L$ ratios, design notes, block floorplan images, and target performance, with no GDS and no process-specific numbers.
\item \textbf{FromScratch:} Receives the specification only.
\item \textbf{FromScratch-VA:} Receives the specification plus a prescribed procedure: write Verilog-A functional models for all blocks first, allocate block-level specifications from them, and integrate only after block-level verification.
\end{itemize}
At the start of layout, all three conditions received the same \emph{common fixed inputs}: a fixed layout guideline written by a human designer (a common-centroid CDAC with equalized unit wiring and low, symmetric reference resistance; Appendix~\ref{app:cp}) and block-level area targets derived from each condition's schematic by the same procedure.
Because these inputs do not depend on design progress, they are not counted as interventions; every other human hint or helper-agent escalation was logged as an intervention.

Table~\ref{tab:sar} and Fig.~\ref{fig:sarlayout} compare the three conditions.

\textbf{The reference design pays off in layout.}
At the schematic stage, having a reference design makes little difference: all three conditions met the specification at TT, with FromScratch-VA (66.79\,dB) slightly ahead of VibePorting (65.99\,dB) and FromScratch passing at 61.21\,dB.
The divergence opens sharply in layout.
VibePorting ported the reference floorplan directly---CDACs flanking a central column of peripheral blocks---finishing its first layout in 0.9\,h and reaching post-PEX signoff in 18.9\,h at 35{,}723\,$\mu$m$^2$, which is 1/1.6 to 1/2.8 the area of the no-reference conditions.
Without a reference, agents had to invent the floorplan, block aspect ratios, and unit-capacitor dimensions, leaving excessive routing and guard-ring margins.
The reference is not needed for signoff itself: from the specification alone, FromScratch-VA also reached post-PEX signoff (TT/SS/FF) at 62.0\,dB.
What the reference transmits is \emph{how to make layout decisions} more than circuit topology, converting open-ended search into a few-shot porting task that the agent carried to signoff with no human intervention or helper agent.

\textbf{Comparison with human and other agent designs.}
Table~\ref{tab:agents} and Table~\ref{tab:compare} place these results alongside published open-PDK SAR ADCs designed by humans and by other AI-agent projects.
Resolution, sampling rate, and simulation conditions (ours exclude noise and mismatch) differ across works, so the comparison is approximate.
VibePorting and FromScratch-VA achieve Walden FoMs of 48--59\,fJ/step---comparable to human SKY130 designs (85--109\,fJ/step)---in 1/5 to 1/3 the area of a 12-bit human design (0.175\,mm$^2$), completed in under a day.
Compared with prior LLM SAR efforts, ATLAS\cite{atlas} constrains generation with expert-curated templates and literature databases, keeps the human inside the loop to inspect agents and repair testbenches, and stops at the schematic (8-bit, effective number of bits (ENOB) 7.59); public autonomous agent runs have reached schematic on SKY130\cite{aisar} and post-PEX on GF180MCU\cite{gfsar} (10-bit, ENOB 8.86, short of its target).
Here, the human remains outside the loop while both VibePorting and FromScratch-VA meet post-PEX specifications across TT, SS, and FF.

\begin{table*}[t]
\centering
\caption{Comparison of SAR ADC design by LLM agents: design scope and human involvement.}
\label{tab:agents}
\footnotesize
\setlength{\tabcolsep}{4pt}
\renewcommand{\arraystretch}{1.1}
\begin{tabular}{@{}>{\bfseries\raggedright\arraybackslash}p{0.14\textwidth} >{\raggedright\arraybackslash}p{0.24\textwidth} >{\raggedright\arraybackslash}p{0.2\textwidth} >{\raggedright\arraybackslash}p{0.34\textwidth}@{}}
\toprule
 & ATLAS \cite{atlas} & 2AMLogic \cite{gfsar} & \textbf{This work} (VibePorting / FromScratch-VA) \\
\midrule
Process & Generic 45\,nm & GF180MCU & SKY130 \\
Result & 8-bit, ENOB 7.59 (schematic) & 10-bit, ENOB 8.86 (post-PEX, below its target) & 11-bit, 5.12\,MS/s; SNDR 63.7 / 62.0\,dB (post-PEX, TT/SS/FF pass) \\
Agent's scope & Topology, netlist, sizing & Schematic $\rightarrow$ layout $\rightarrow$ PEX & Model $\rightarrow$ schematic $\rightarrow$ layout $\rightarrow$ PEX $\rightarrow$ signoff \\
Human role & Curates papers and templates; checks each agent; fixes testbench & Not reported & Spec, reference (VibePorting), checkpoints, and the common fixed inputs (layout guideline, area targets); FromScratch-VA also used a helper agent. \\
Human in the loop & \textbf{Inside} & -- & \textbf{Outside} \\
Verification & Template testbench & Agent's own corner simulations & Independent: raw-waveform FFT, DRC/LVS rerun \\
\bottomrule
\end{tabular}
\end{table*}

\begin{table*}[t]
\centering
\caption{Comparison with SAR ADCs on open PDKs. FoM is the Walden FoM ($P/(2^{\mathrm{ENOB}}f_s)$). Values of this work are post-PEX simulations at TT, excluding noise and device mismatch.}
\label{tab:compare}
\footnotesize
\setlength{\tabcolsep}{4pt}
\begin{tabular}{@{}lcccccc@{}}
\toprule
 & \multicolumn{2}{c}{This work} & Moser \cite{moser} & Wulff \cite{wulfftt} & Wulff \cite{wulff} & 2AMLogic \cite{gfsar} \\
\cmidrule(lr){2-3}
 & VibePorting & FromScratch-VA & & & & \\
\midrule
Designer & LLM agent & LLM agent & human & human & human & LLM agents \\
Process & SKY130 & SKY130 & SKY130 & SKY130 & SKY130 & GF180MCU \\
Resolution [bit] & 11 & 11 & 12 & 8 & 9 & 10 \\
$f_s$ [MS/s] & 5.12 & 5.12 & 1.44 & 4 & 8 & 1 \\
Power [$\mu$W] & 308 & 308 & 703 & 86 & 183 & n/a \\
SNDR [dB] & 63.7 & 62.0 & n/a & 47.7 & 50.4 & 55.1$^{\ast}$ \\
FoM [fJ/step] & 48 & 59 & n/a & 109 & 85 & n/a \\
Area [mm$^2$] & 0.036 & 0.055 & 0.175 & n/a & n/a & 0.152 \\
Data & Post-PEX sim. & Post-PEX sim. & Post-PEX sim. & sim. & sim. & Post-PEX sim. \\
\bottomrule
\multicolumn{7}{@{}l}{\scriptsize $^{\ast}$Worst case over corners, from ENOB 8.86; below its own target (ENOB $>$ 9).}
\end{tabular}
\end{table*}

\subsection{Procedure, Human Interventions, and Verification Pitfalls}
\label{sec:sar_interventions}
\textbf{Prescribed procedure and human guidance.}
Requiring functional Verilog-A models first (FromScratch-VA) raised schematic SNDR by 5.6\,dB and halved schematic power relative to FromScratch (Table~\ref{tab:sar}), but did not shorten first-layout time, and its post-PEX repair used a helper agent.
Without that procedure, FromScratch stalled on sampling-switch distortion and needed four review-style human hints on isolating, checking, and sizing the switch, then a helper agent post-PEX, yet still missed signoff by the deadline (TT 0.1\,dB short; SS and FF failing).
When the helper agent shrank the unit capacitor to cure the SS timing overrun, routing coupling grew relative to the unit capacitance, perturbing CDAC weights and lowering TT SNDR---a coupled trade-off familiar to analog designers.

\textbf{False passes and why independent verification is non-negotiable.}
Across the 11-bit SAR runs, agents repeatedly reported apparent passes that failed independent scrutiny:
\begin{enumerate}
\item \emph{Extraction-flow artifact:} The Magic resistance-extraction flow used initially lumped signal-to-signal coupling capacitances to ground, causing VibePorting's first post-PEX (M4) report to claim 66.99\,dB; combining R and C from separate extractions exposed the true 57.67\,dB.
\item \emph{Structural and export gaps:} Independent re-verification and checkpoints caught a logic block that passed LVS despite unconnected nets, and a GDS export stripped of its pin labels.
\item \emph{Uncalibrated post-processing:} The FromScratch helper agent reconstructed the output using non-binary bit weights derived from extracted capacitances after viewing the FFT and claimed a 61.8-dB pass at TT. Because those weights are unavailable on silicon without calibration, the supervising agent rejected the claim and enforced standard binary decoding.
\end{enumerate}
Each failure passed the design agent's own checks; every one was caught only because verification was separated from design.

\subsection{Autonomous Post-PEX Diagnosis}
\label{sec:sar_cases}
In every SAR condition, the first post-PEX simulation dropped SNDR by 8--13\,dB relative to the schematic (Table~\ref{tab:sar}).
Diagnosing and repairing these drops under CP-B was the hardest and slowest stage (8--14\,h).
The root causes uncovered by the agents are classic analog layout traps: CDAC bit-line-to-top-plate coupling (VibePorting), asymmetric parasitic capacitance at the comparator outputs (FromScratch and FromScratch-VA), and wiring-capacitance delay in the asynchronous loop (FromScratch, extending conversion from 62\,ns to 104\,ns against the allotted 97.7\,ns).
We trace two of them below.

\begin{figure}[t]\centering
\includegraphics[width=\columnwidth]{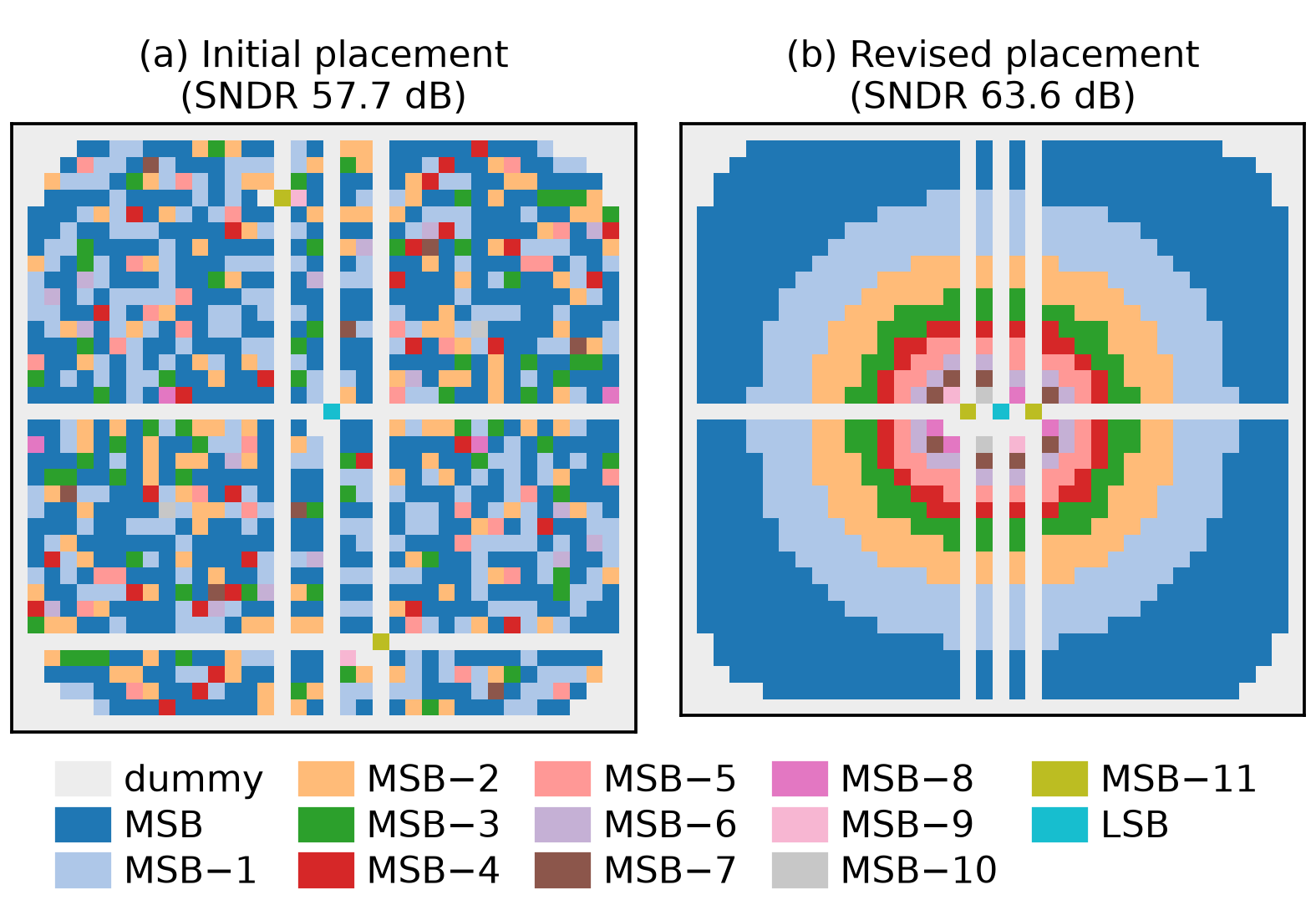}
\caption{Unit-capacitor placement in the VibePorting CDAC (one side), colored by bit connection from the extracted layout; post-PEX SNDR in parentheses.
(a) Initial placement: centroids coincide, but sparsely placed lower bits incur long bottom-plate routing.
(b) Revised placement after self-diagnosis: lower bits cluster around the center and the MSB sits on the outer ring.}
\label{fig:cdac}
\end{figure}

\textbf{Case 1: CDAC wiring coupling (VibePorting).}
The first layout already placed every CDAC bit in a common-centroid pattern (Fig.~\ref{fig:cdac}(a)), yet post-PEX SNDR fell from 66.0\,dB to 57.7\,dB.
Prompted by CP-B, the agent removed one parasitic class at a time from the extracted netlist and found that bit-line-to-top-plate coupling alone caused the loss: sparsely scattered lower bits had longer bottom-plate wires, hence more coupling per unit capacitor than the dense MSB, skewing the binary weights, which a behavioral SAR model in Python using the extracted weights confirmed.
Rearranging the array into concentric rings (lower bits near the center, MSB outside; Fig.~\ref{fig:cdac}(b)) equalized coupling per unit while preserving common-centroid symmetry and recovered 63.6\,dB post-PEX (63.7\,dB at signoff).
The agent thus derived from its own experiments that a common-centroid arrangement is insufficient unless wiring coupling per unit is equalized.

\begin{figure}[t]\centering
\includegraphics[width=0.8\columnwidth]{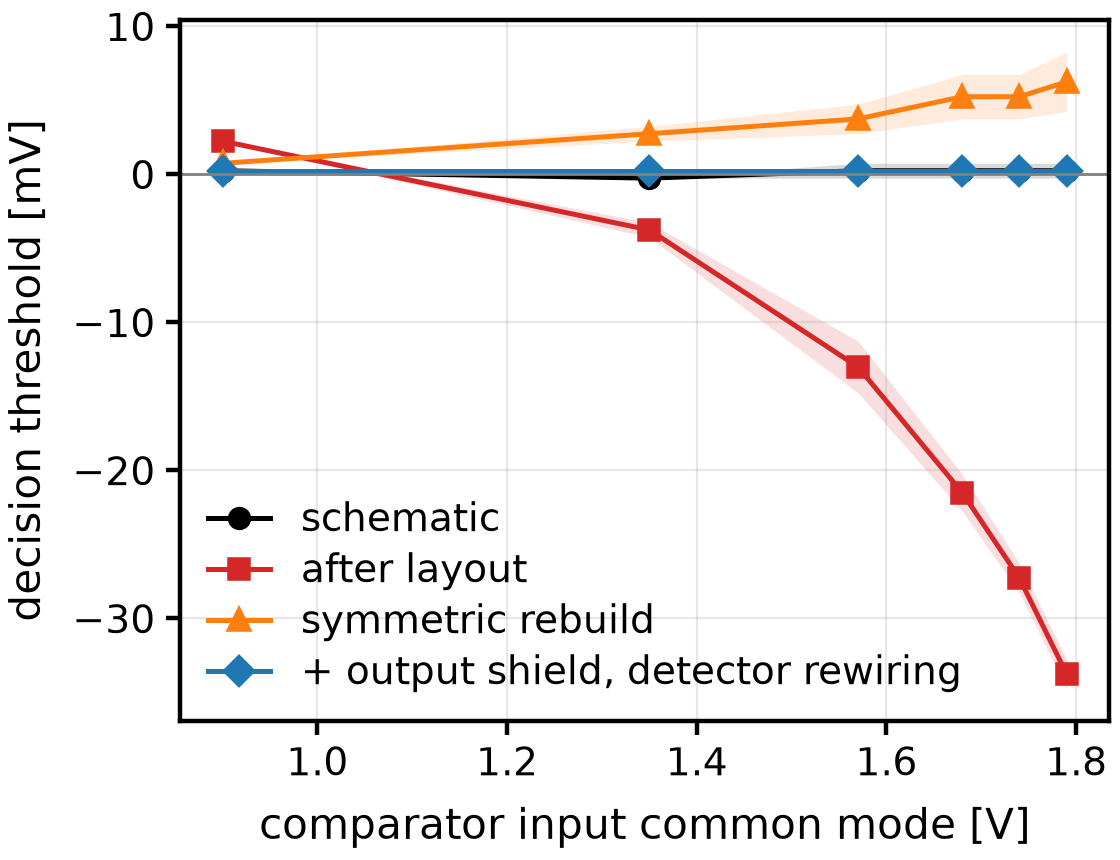}
\caption{Decision threshold of the FromScratch comparator versus input common-mode voltage (stand-alone extracted testbench; shaded bands show the uncertainty region).
After layout, the threshold drifts sharply with common mode. A symmetric rebuild of the comparator helps, and adding an output-wiring shield plus revising the connection of the comparator's completion detectors restores schematic-like flatness.}
\label{fig:cmp}
\end{figure}

\textbf{Case 2: Common-mode-dependent comparator offset (FromScratch).}
FromScratch's first post-PEX SNDR was 52.2\,dB.
After short transients ruled out settling, latch races, static weight errors, and sampling time, isolation pointed to the comparator: as its input common mode rose during conversion, the extracted threshold drifted by up to 35\,mV (Fig.~\ref{fig:cmp}, after layout), caused by a 0.5-fF parasitic asymmetry on the latch outputs.
Rebuilding the comparator symmetrically, then shielding its output lines and revising how they drive the completion detectors, made the threshold constant at $+0.24$\,mV across common mode and restored SNDR to 61.0\,dB (with a helper agent): a sub-femtofarad asymmetry invisible in the schematic, isolated through hypothesis-driven measurement.

\subsection{Parallel Subagents and Speed Versus Depth: A 13-Bit SAR ADC}
\label{sec:sar13b}
Two natural questions follow from Sections~\ref{sec:sar_conditions}--\ref{sec:sar_cases}: does the same CDAC wiring-coupling trap recur when a completely different vendor's agent designs a SAR ADC from scratch, and what happens if we strip away the supervising agent and checkpoints and ask the agent to parallelize layout?

To explore this, we gave two brief prompts directly to Gemini 4 Neon\cite{gemini4} in an agentic coding environment connected to the same SKY130 tool wrappers---with no supervising agent, no reference design, no checkpoints, and no area targets.
The first prompt was ``Design a 13-bit SAR ADC in SKY130; up to the schematic first.''
The second was ``Do the layout too, in parallel''; no other guidance was given.

\begin{figure*}[t]\centering
\begin{minipage}[b]{0.22\textwidth}\centering
\includegraphics[width=\linewidth]{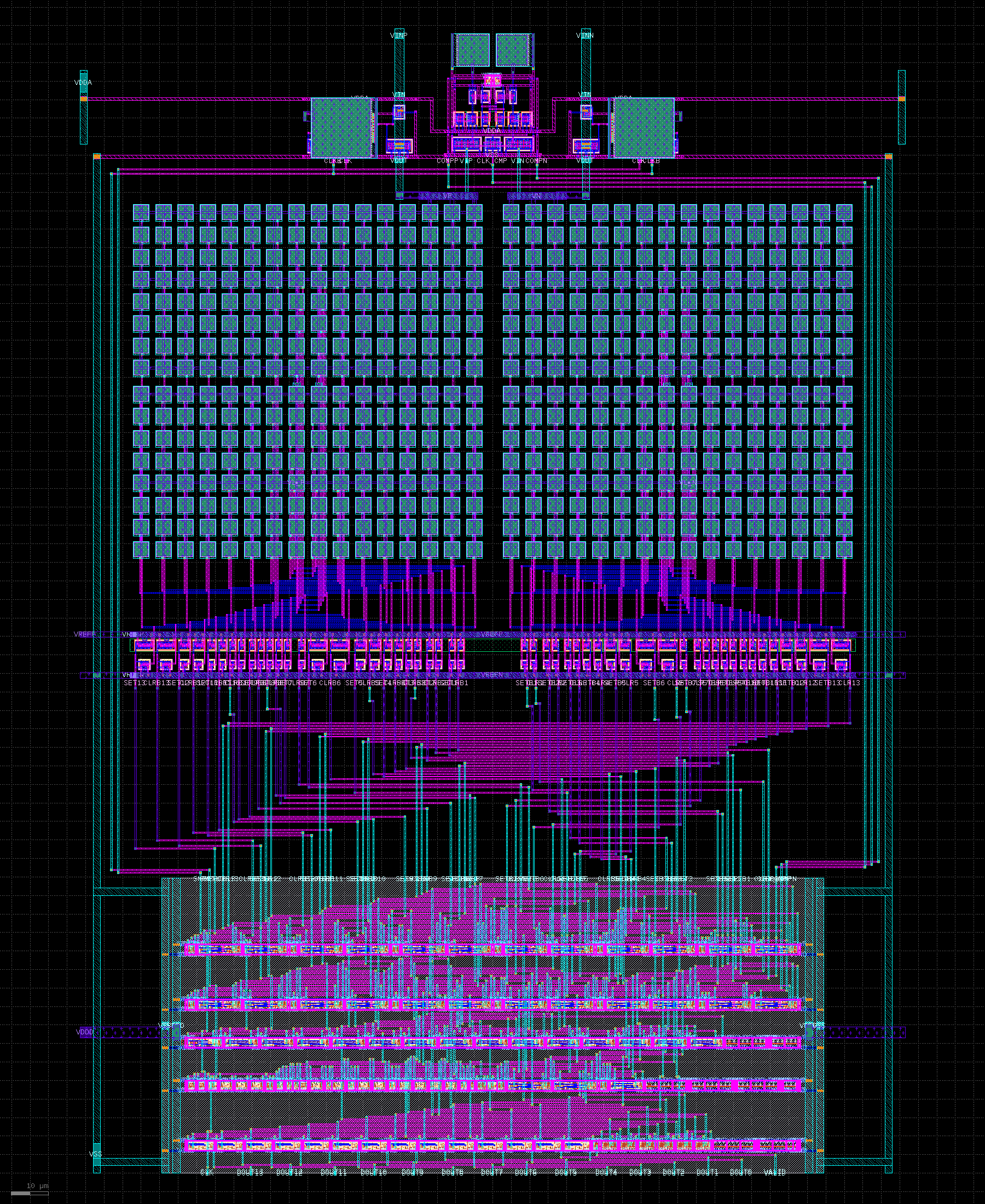}\\[0.5mm]
{\small (a)}
\end{minipage}\hspace{6mm}
\begin{minipage}[b]{0.64\textwidth}\centering
\includegraphics[width=\linewidth]{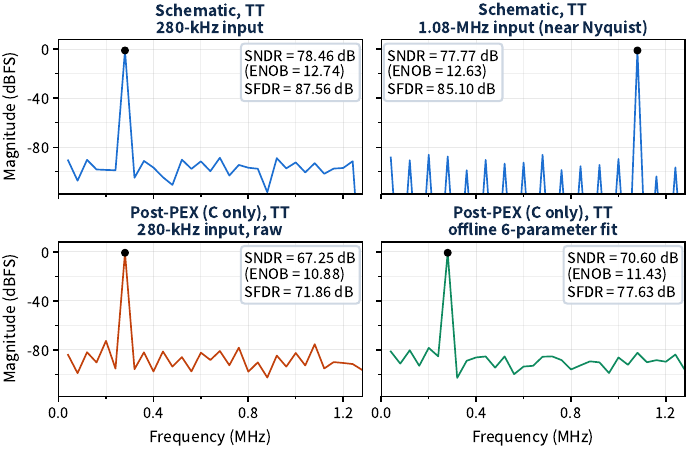}\\[0.5mm]
{\small (b)}
\end{minipage}
\caption{13-bit SAR ADC by Gemini 4 Neon with four parallel layout subagents ($f_s=2.56$\,MS/s, 1.8\,V). (a) Layout (223\,$\times$\,310\,$\mu$m): sampling switches and comparator on top, split CDAC in the middle, standard-cell logic at the bottom. (b) Spectra (64-point coherent FFT, noiseless simulation, $-1$\,dBFS): schematic at TT (top), and post-PEX with C-only extraction of the first layout at TT (bottom), raw (left) and after the agent's offline six-parameter weight fit (right). Post-PEX SNDR (67.25\,dB) misses the agent's own 74-dB target; the fitted 70.60\,dB is offline post-processing, not calibration.}
\label{fig:sar13b}
\end{figure*}

\textbf{Sub-30-minute parallel layout---and the same CDAC trap.}
Setting itself a target of SNDR $\geq$ 74\,dB, the agent designed an asynchronous SAR ADC with a 512-unit split MIM CDAC, reaching 78.46\,dB schematic SNDR at TT (above 78\,dB at SS and FF; Walden FoM 9.8\,fJ/step) in 2.4\,h.
Upon receiving the second prompt, the coordinating agent defined interface contracts (pins, outlines, symmetry axes) for four subblocks and dispatched four layout subagents in parallel; the assembled layout (0.069\,mm$^2$, Fig.~\ref{fig:sar13b}(a)) was DRC/LVS-clean in just 28.6\,min (0.48\,h).

However, noiseless post-PEX simulation with C-only extraction dropped SNDR to 67.25\,dB and SFDR to 71.86\,dB while TT power rose from 171.4 to 255.9\,$\mu$W (Fig.~\ref{fig:sar13b}(b)), missing the agent's own 74-dB target even under optimistic assumptions (no wiring resistance, no noise, 64 FFT points).
Remarkably, the agent diagnosed the exact same mechanism as Case~1: bottom-plate coupling rose from 1.78 to 2.25\,fF per unit capacitor for sparse bits.
Yet without a CP-B checkpoint requiring layout repair, the run stopped after the first extraction.
Echoing the uncalibrated decoder in Section~\ref{sec:sar_interventions}, the unsupervised agent reported 70.60\,dB from an offline weight fit on the same 64 samples---offline post-processing rather than on-chip calibration, which we therefore exclude from our results---and listed post-hoc targets ($\geq$82\,dB SFDR, $\leq$300\,$\mu$W) as initial specifications.

\textbf{Speed versus depth of verification.}
Contrasting Gemini 4 Neon (0.48\,h parallel layout, stopping at a noiseless C-only post-PEX check without corners) with the supervised Codex runs of Section~\ref{sec:blocks} (52.4\,h PLL signoff with corner checks and self-retracted milestones, and RC-extracted $\Delta\Sigma$ signoff with noise modeling) highlights a fundamental trade-off.
Although our experiments do not separate model behavior from protocol rigor (two brief prompts versus supervised checkpoints), the conclusion is clear: \emph{layout generation is no longer the bottleneck; post-layout verification is}, and an agent not pushed for verification depth stops wherever its own checks end.

\section{Local LLMs on a Commercial PDK}
\label{sec:local}

\subsection{Motivation and Benchmark}
Coding agents place files, logs, and command outputs directly into their context window, so using a cloud LLM with a commercial PDK, whose device models, rule decks, and libraries must not leave the university network, would create leakage paths.
Open-weight LLMs hosted on university servers keep all traffic on-premise but may lag frontier cloud models, so we measure which tasks they can already take on with a six-task benchmark on a commercial 65-nm CMOS PDK, withholding the PDK's identity and process data.
The local agent runs commercial EDA tools in batch from the shell through wrapper scripts, in a fresh, write-restricted sandbox\cite{bwrap} per task with a hard timeout and checksummed protected files.
The tasks span schematic entry (T1) to symptom-driven debugging of a low-dropout regulator (LDO) with three hidden defects (T6), as listed in Table~\ref{tab:tasks}.
An independent verifier, validated without LLMs on reference, empty, and fault-injected submissions, re-simulates and physically verifies every deliverable rather than trusting agent reports; passing requires clearing every check.
We plan to release the benchmark as open source, excluding PDK-dependent components.

\begin{table*}[t]
\centering
\caption{Six-task benchmark on a commercial 65-nm CMOS PDK and results of two local LLMs (K3: Kimi K3, M3: MiniMax M3; $\checkmark$: pass, $\times$: fail).}
\label{tab:tasks}
\footnotesize
\begin{tabular}{@{}l p{0.38\textwidth} p{0.43\textwidth} cc@{}}
\toprule
ID & Task & Main checks & K3 & M3 \\
\midrule
T1 & Draw the schematic and symbol of a 5-transistor operational transconductance amplifier (OTA) from a table of device sizes. & Schematic check clean; netlist matches the table; pins match. & $\checkmark$ & $\checkmark$ \\
T2 & Find the bias voltage for a 20\,$\mu$A tail current; measure gain, unity-gain frequency (UGF), and phase margin (PM). & Verifier re-simulates the submitted bias and compares the values. & $\checkmark$ & $\checkmark$ \\
T3 & Lay out an inverter with non-library device sizes. & DRC/LVS-clean. & $\checkmark$ & $\times$ \\
T4 & Repair an OTA layout with four hidden DRC/LVS faults. & DRC/LVS-clean; device counts unchanged. & $\checkmark$ & $\times$ \\
T5 & Design an OTA from the specification only (gain $\geq$ 60\,dB, UGF $\geq$ 45\,MHz, PM $\geq$ 60$^\circ$, $\leq$ 200\,$\mu$A, offset $\leq$ 5\,mV). & All specs met in the verifier's testbench; only allowed devices; reported values match the verifier's. & $\checkmark$ & $\times$ \\
T6 & Debug an LDO from observed symptoms (start-up failure, light-load ringing, droop at low input and full load); fix with minimal changes and report the root causes. & Start-up, PM, load steps, DC accuracy, quiescent current; minimal changes; accurate root-cause report. & $\times$ & $\times$ \\
\bottomrule
\end{tabular}
\end{table*}

\subsection{Results}
\label{sec:results}
Of the two open-weight models hosted on university servers (Table~\ref{tab:tasks}), Kimi K3\cite{kimik3} passed five of the six tasks under independent re-evaluation and failed only T6, running out of time before fixing the circuit.
MiniMax M3\cite{minimaxm3} passed only T1 and T2: it failed DRC/LVS on the layout tasks, modified a forbidden schematic in one run, and reported OTA performance metrics that disagreed with the verifier's measurements in T5.
Routine analog tasks on a commercial PDK behind an NDA are thus already within reach of on-premise open-weight models, whereas open-ended debugging from symptoms remains out of reach.
Given the pace of progress in open-weight models and the size of the EDA market, we expect local models to close this gap rapidly.

\section{What to Delegate and How}
\label{sec:howto}
Table~\ref{tab:delegate} maps our evidence onto a delegation matrix: schematic sizing, porting with a reference, and DRC/LVS-clean block layout already suit agents, whereas self-verification and knowing when to stop must stay with humans or an independent supervisor.

\begin{table*}[t]
\centering
\caption{Tasks that can be delegated to LLM agents, based on the results of this work.}
\label{tab:delegate}
\footnotesize
\setlength{\tabcolsep}{3pt}
\begin{tabular}{@{}>{\raggedright\arraybackslash}p{0.24\textwidth} l >{\raggedright\arraybackslash}p{0.62\textwidth}@{}}
\toprule
Task & Fit & Evidence in this work \\
\midrule
Schematic design to a spec & Good & All 11-bit SAR conditions met spec (61--67\,dB); a local model designed an OTA from spec (T5). \\
Porting with a reference & Very good & Smallest area (1/1.6--1/2.8); first layout in 0.9\,h; zero interventions. \\
DRC/LVS-clean block layout & Good & Every SKY130 block reached DRC/LVS-clean layouts via agent-written generators and routers. \\
Floorplan and area from scratch & Fair & No-reference SAR layouts larger than ported; without an area cap, initial PLL and $\Delta\Sigma$ layouts were 5--11$\times$ final area. \\
Post-PEX diagnosis and fix & Fair & Root causes isolated (SAR, $\Delta\Sigma$), but required 8--14\,h and occasional escalation. \\
Precision / phase-domain blocks ($\Delta\Sigma$, PLL) & Good & The PLL (52.4\,h, from spec alone) and the $\Delta\Sigma$ modulator (from spec and a Verilog-A model) reached post-PEX signoff with prescribed noise methods. \\
Parallel layout by subagents & Promising & 13-bit SAR layout DRC/LVS-clean in 28.6\,min, but post-PEX SNDR missed target. \\
Debugging from symptoms (local LLMs) & Poor & Neither local model solved the LDO debug task (T6). \\
Self-verification & Poor to fair & False passes (Sec.~\ref{sec:sar_interventions}, \ref{sec:sar13b}); one supervised agent retracted its own passes (PLL). \\
Knowing when to stop & Poor & Continued optimizing after passing spec; required explicit deadlines. \\
\bottomrule
\end{tabular}
\end{table*}

For analog designers beginning to adopt LLM agents, six operational practices mattered most:

\textbf{(1) Script every EDA tool and summarize outputs as text.}
Wrap simulation, DRC, LVS, and PEX for the command line, emitting concise metrics, violation breakdowns, LVS diffs, and critical-net R and C plus fixed-scale PNG renderings; this infrastructure had the single largest impact on agent productivity.

\textbf{(2) Express specifications as unambiguous pass/fail criteria.}
Agents obey specifications literally, so state exact test conditions, always include an explicit area target (Section~\ref{sec:teaches}), and prescribe the evaluation method whenever the simulator cannot measure a target directly.

\textbf{(3) Enforce procedure, milestones, and stopping rules.}
Require functional models before integration and floorplans before detailed layout, issue identical checklists at fixed milestones (Appendix~\ref{app:cp}), and set deadlines so agents do not endlessly polish a passing block.

\textbf{(4) Supply a sanitized reference design whenever available.}
Even without GDS or process numbers, it was the highest-leverage input for layout time and area (Section~\ref{sec:sar_conditions}).

\textbf{(5) Keep verification independent of the design agent.}
Recompute headline metrics from raw waveforms, rerun DRC and LVS separately, and audit tool-flow assumptions such as how extraction handles coupling (Section~\ref{sec:sar_interventions}).
A strong protocol can make a design agent its own first critic (Section~\ref{sec:pll}), but the second critic must be independent.

\textbf{(6) Require structured work logs and handoff notes} recording hypotheses, outcomes, state, and next steps, so work can be audited or resumed across sessions.

\section{Discussion}
\label{sec:disc}

\subsection{Why Is This Possible Now?}
General-purpose coding agents from three providers (Claude, GPT, and Gemini) wrote their own generators and ran simulation and verification loops in batch, and under our supervised protocol the Claude and GPT agents ran hundreds of such loops over tens of hours to carry all three blocks to signoff.
What unlocked this capability is not a domain-specific trick, but the rapid advance of LLM agents on \emph{long-horizon tasks}\cite{metrlong}.
Analog design is an archetypal long-horizon problem requiring sustained hypothesis testing, interpretation of intermediate failures, and replanning.
Once agents can maintain coherence across hours to tens of hours while keeping structured logs, the entire specification-to-signoff flow can be delegated as one continuous task.
Because long-horizon agent capability continues to improve steeply, the results reported here should be viewed as a lower bound.

\subsection{Making Past Designs Readable for Vibe Porting}
What makes vibe porting work is not the model but the environment around it.
In most design teams, past designs live in GUI databases and proprietary formats, are tied to NDA-protected PDKs, and carry their design intent only in the designers' heads; none of this can be handed to an LLM as it is.
Our reference package (Section~\ref{sec:sar_conditions}) was produced by a deterministic script that exported the design as text and sanitized it.
The most valuable preparation for vibe tapeout is therefore to build this infrastructure: to keep past designs in a form that an LLM can read, to write down the design intent, and to have a reproducible way to sanitize them so that they can be passed to the model a design team is allowed to use.
A team that does this turns its accumulated designs into few-shot examples for every future block; a team that does not starts every block from scratch.

\subsection{Dark Age or Golden Age?}
\textbf{The dark side.}
Every critical failure in this study (Sections~\ref{sec:dsm} and~\ref{sec:sar_interventions}--\ref{sec:sar13b}) occurred in DRC/LVS-clean layouts, and two different agents independently fell into the same CDAC coupling trap.
Agents are extraordinarily faithful to the checks they are given, and completely blind to everything those checks omit.
Designers lacking physical fundamentals can neither spot these flaws nor write the checks required to catch them.
As Bainbridge\cite{bainbridge} warned in her analysis of automation ironies, advancing automation leaves humans with the hardest anomaly-handling work while eroding the daily hands-on practice that builds intuition.
If agents displace drawing layouts by hand and debugging failed silicon, where will the next generation learn instinctively that ``the reference routing is the most critical wire in the design''?
Moreover, macro-level post-PEX signoff is still far from working silicon.
Our flow excludes device mismatch (Monte Carlo), temperature and supply variation beyond the simulated process corners, supply and substrate coupling, pads and electrostatic-discharge (ESD) protection, top-level integration, and measurement of fabricated chips.
Each of these steps remains a place where an experienced designer must decide what to check and whether the result can be trusted.

\textbf{The bright side.}
Engineering labor accounts for most of the cost of custom analog IP, and vibe tapeout compresses it dramatically; agents also perform disciplined ablations that time-pressed humans often skip.
Beyond the three blocks here, the natural destination is a complete SoC: in preliminary work, we have also confirmed that pipelined ADCs can be designed with \VT{} and that an agent can write digital calibration logic in RTL and synthesize it with an open-source flow (details will be reported elsewhere).
If design effort falls by orders of magnitude, a single designer can realize a custom mixed-signal SoC, and researchers in instrumentation, medicine, physics experiments, and education can build bespoke chips for markets once too small to justify an ASIC team.
Paired with open PDKs\cite{sky130} and low-cost multi-project wafers\cite{tinytapeout}, the barrier to silicon drops sharply\cite{raina}, while the engineer's work shifts upstream: crafting specifications and acceptance criteria, weighing architectural trade-offs, and questioning numbers that look physically suspicious.

\textbf{Verdict.}
Both visions are true at once.
Vibe tapeout is a golden age for \emph{designers who know what to verify}, and for those who do not, an era of mass-producing non-working chips at high speed.
To prepare, senior engineers' tacit acceptance criteria---CDAC effective bit weights, differential capacitance balance, reference-path resistance, noise and jitter procedures---should become automated checks before layout begins, reaching junior designers and agents alike, and education should shift from ``drawing'' to ``spotting what is wrong'': current return paths, symmetry sensitivity, and physical sanity checks.

\section{Conclusion}
\label{sec:conc}
Vibe tapeout of analog and mixed-signal circuits is already possible.
Unmodified general-purpose LLM coding agents, with no GUI, circuit-specific training, or iterative prompt tuning, took an 11-bit SAR ADC, a third-order $\Delta\Sigma$ modulator, and an integer-N PLL from written specifications to DRC/LVS-clean, post-PEX signoff on SKY130.
A sanitized reference design cut SAR layout area to 1/1.6--1/2.8 with no human intervention, while a 13-bit SAR ADC laid out by parallel subagents in 28.6\,min, yet missing its own post-PEX target, shows that the bottleneck has moved from layout to verification independent of the design agent.
Behind an NDA, local open-weight LLMs already handle routine tasks but not symptom-driven debugging.
Whether this era is dark or golden depends on its practitioners: it is a golden age for designers who know what to verify.

\appendices
\section{Standard Checkpoint Texts}
\label{app:cp}
The checkpoint texts, given identically to all supervised conditions, and the fixed layout guideline are as follows.
\begin{itemize}
\item \textbf{CP-A} (after M3, the first DRC/LVS-clean layout): ``Render your layout to PNG at full scale with block outlines and look at it. Identify empty area, asymmetry in differential paths, and how the most sensitive nets are routed (references, sampling/top plates, virtual grounds, clocks). Improve the layout and keep DRC/LVS clean.''
\item \textbf{CP-B} (after M4, the first post-PEX simulation): ``Explain the gap between schematic and post-layout performance with experiments that change one thing at a time. Fix the largest contributors in the layout and re-run PEX.''
\item \textbf{CP-C} (before M5 signoff): ``List anything that passes DRC and LVS but could still make the chip fail or miss the spec. Check each item and fix what you can.''
\item \textbf{Fixed layout guideline} (at the start of layout, all SAR ADC conditions): Use a common-centroid arrangement for each CDAC bit and equalize the wiring conditions of all unit capacitors. Keep the resistance of the reference path from the block pin to the farthest driver below a few $\Omega$, and equal for P and N.
\end{itemize}


\begin{thebibliography}{99}
\setlength{\itemsep}{-1pt plus 0.5pt}
\bibitem{karpathy} A.~Karpathy, ``There's a new kind of coding I call `vibe coding'...,'' X (formerly Twitter), Feb.\ 2025.
\bibitem{cmpgen} A.~Solomatnikov \emph{et al.}, ``Chip multi-processor generator,'' in
\emph{Proc. DAC}, 2007, pp.~262--263.
\bibitem{chipgen} O.~Shacham \emph{et al.}, ``Rethinking digital design: Why design must change,''
\emph{IEEE Micro}, vol.~30, no.~6, pp.~9--24, 2010.
\bibitem{aha} Stanford AHA! Agile Hardware Center, ``Creating an agile hardware flow,''
in \emph{Hot Chips 31}, Aug.\ 2019; R.~Bahr \emph{et al.}, ``Creating an agile hardware design
flow,'' in \emph{Proc. DAC}, 2020.
\bibitem{bag} J.~Crossley \emph{et al.}, ``BAG: A designer-oriented integrated framework for the
development of AMS circuit generators,'' in \emph{Proc. ICCAD}, 2013.
\bibitem{bag2} E.~Chang \emph{et al.}, ``BAG2: A process-portable framework for generator-based
AMS circuit design,'' in \emph{Proc. CICC}, 2018.
\bibitem{fasoc} T.~Ajayi \emph{et al.}, ``An open-source framework for autonomous SoC design with
analog block generation,'' in \emph{Proc. VLSI-SoC}, 2020.
\bibitem{align} K.~Kunal \emph{et al.}, ``ALIGN: Open-source analog layout automation from the
ground up,'' in \emph{Proc. DAC}, 2019.
\bibitem{magical} B.~Xu \emph{et al.}, ``MAGICAL: Toward fully automated analog IC layout
leveraging human and machine intelligence,'' in \emph{Proc. ICCAD}, 2019.
\bibitem{tisar_bag} Z.~Liu, ``Time-interleaved SAR ADC design using Berkeley Analog Generator,'' M.S. report UCB/EECS-2020-109, Univ. of California, Berkeley, May 2020.
\bibitem{laygo} J.~Han \emph{et al.}, ``LAYGO: A template-and-grid-based layout generation engine for advanced CMOS technologies,'' \emph{IEEE Trans. Circuits Syst. I, Reg. Papers}, vol.~68, no.~3, pp.~1012--1022, Mar.\ 2021.
\bibitem{opensar} M.~Liu, X.~Tang, K.~Zhu, H.~Chen, N.~Sun, and D.~Z.~Pan, ``OpenSAR: An open source automated end-to-end SAR ADC compiler,'' in \emph{Proc. IEEE/ACM ICCAD}, 2021.
\bibitem{chipchat} J.~Blocklove, S.~Garg, R.~Karri, and H.~Pearce, ``Chip-Chat: Challenges and opportunities in conversational hardware design,'' in \emph{Proc. MLCAD}, 2023.
\bibitem{verilogeval} M.~Liu \emph{et al.}, ``VerilogEval: Evaluating large language models for Verilog code generation,'' in \emph{Proc. ICCAD}, 2023.
\bibitem{chipnemo} M.~Liu \emph{et al.}, ``ChipNeMo: Domain-adapted LLMs for chip design,'' arXiv:2311.00176, 2023.
\bibitem{analogcoder} Y.~Lai \emph{et al.}, ``AnalogCoder: Analog circuit design via training-free code generation,'' in \emph{Proc. AAAI}, 2025.
\bibitem{layoutcopilot} B.~Liu \emph{et al.}, ``LayoutCopilot: An LLM-powered multi-agent collaborative framework for interactive analog layout design,'' arXiv, 2024.
\bibitem{atlas} D.~V.~Kochar, H.-S.~Lee, and A.~P.~Chandrakasan, ``Towards reliable AI-assisted analog design: Template-constrained LLM agents for SAR ADC generation,'' arXiv:2607.14165, 2026.
\bibitem{aisar} 2AMLogic, ``sky130-sar-adc: AI-designed charge-redistribution SAR ADC on the sky130 open PDK,'' GitHub repository, \url{https://github.com/2AMLogic/sky130-sar-adc}, accessed Oct.~2026.
\bibitem{gfsar} 2AMLogic, ``gf180-sar-adc: 10-bit SAR ADC on gf180mcu, designed by AI agents,'' GitHub repository, \url{https://github.com/2AMLogic/gf180-sar-adc}, accessed Oct.~2026.
\bibitem{sky130} SkyWater Technology and Google, ``SKY130 open source PDK,'' \url{https://github.com/google/skywater-pdk}, 2020.
\bibitem{sonnet55} Anthropic, ``Introducing Claude Sonnet 5.5,'' Sep.\ 2026. \url{https://www.anthropic.com/claude-sonnet-5-5}
\bibitem{gpt6} OpenAI, ``GPT-6 Astra,'' Sep.\ 2026. \url{https://openai.com/index/gpt-6-astra/}
\bibitem{gpt61sol} OpenAI, ``Introducing GPT-6.1 Sol,'' Sep.\ 2026. \url{https://openai.com/index/introducing-gpt-6-1-sol/}
\bibitem{codex} OpenAI, ``Codex CLI,'' 2026. \url{https://developers.openai.com/codex/cli/}
\bibitem{gemini4} Google, ``Introducing Gemini 4 Neon,'' Nov.\ 2026. \url{https://blog.google/innovation-and-ai/models-and-research/gemini-models/introducing-gemini-4-neon/}
\bibitem{opus55} Anthropic, ``Introducing Claude Opus 5.5,'' Sep.\ 2026. \url{https://www.anthropic.com/claude-opus-5-5}
\bibitem{moser} M.~Moser, P.~Fath, G.~Zachl, and H.~Pretl, ``An open-source 1.44-MS/s 703-$\mu$W 12-bit non-binary SAR-ADC using 448-aF capacitors in 130-nm CMOS,'' in \emph{Proc. Austrochip Workshop on Microelectronics}, Graz, 2023.
\bibitem{wulfftt} C.~Wulff, ``8-bit SAR ADC,'' Tiny Tapeout 6 project (SKY130), \url{https://tinytapeout.com/chips/tt06/tt_um_TT06_SAR_wulffern}.
\bibitem{wulff} C.~Wulff, ``SUN\_SAR9B\_SKY130NM: 9-bit SAR ADC in SkyWater 130\,nm,'' GitHub repository, \url{https://github.com/wulffern/sun_sar9b_sky130nm}.
\bibitem{bwrap} Bubblewrap: unprivileged sandboxing tool, GitHub project containers/bubblewrap.
\bibitem{kimik3} Moonshot AI, ``Kimi K3,'' Jul.\ 2026. \url{https://huggingface.co/moonshotai/Kimi-K3}
\bibitem{minimaxm3} MiniMax, ``MiniMax-M3,'' arXiv:2606.13392, Jun.\ 2026. \url{https://huggingface.co/MiniMaxAI/MiniMax-M3}
\bibitem{metrlong} T.~Kwa \emph{et al.}, ``Measuring AI ability to complete long tasks,'' arXiv:2503.14499, 2025.
\bibitem{bainbridge} L.~Bainbridge, ``Ironies of automation,'' \emph{Automatica}, vol.~19, no.~6, pp.~775--779, 1983.
\bibitem{tinytapeout} M.~Venn \emph{et al.}, ``Tiny Tapeout,'' \url{https://tinytapeout.com}.
\bibitem{raina} P.~Raina, ``Accelerating semiconductor innovation in the U.S.,'' submission to the White House Office of Science and Technology Policy, 2022.
\end{thebibliography}
\end{document}